# Efficient Design of Helical Higher-Order Topological Insulators in 3D Elastic Medium

Jiachen Luo[1], Zongliang Du[1,2*], Hui Chen[3], Xianggui Ding[1], Chang Liu[1,2], Weisheng Zhang[1,2], Xu Guo[1,2*]

[1]State Key Laboratory of Structural Analysis for Industrial Equipment, Department of Engineering Mechanics, Dalian University of Technology, Dalian, 116023, China
[2]Ningbo Institute of Dalian University of Technology, Ningbo, 315016, China
[3]Piezoelectric Device Laboratory, School of Mechanical Engineering and Mechanics, Ningbo University, Ningbo 315211, China

E-mail: zldu@dlut.edu.cn (ZD); guoxu@dlut.edu.cn (XG)

## Abstract

Topological materials (TMs) are well-known for their topological protected properties. Phononic system has the advantage of direct observation and engineering of topological phenomena on the macroscopic scale. For the inverse design of 3D TMs in continuum medium, however, it would be extremely difficult to classify the topological properties, tackle the computational complexity, and search solutions in an infinite parameter space. This work proposed a systematic design framework for the 3D mechanical higher-order topological insulators (HOTIs) by combining the symmetry indicators (SI) method and the moving morphable components (MMC) method. The 3D unit cells are described by the MMC method with only tens of design variables. By evaluating the inherent singularity properties in the 3D mechanical system, the classic formulas of topological invariants are modified accordingly for elastic waves. Then a mathematical formulation is proposed for designing the helical multipole topological insulators (MTIs) featured corner states and helical energy fluxes, by constraining the corresponding topological invariants and maximizing the width of band gap. Mechanical helical HOTIs with different symmetries are obtained by this method and verified by full wave simulations. This design paradigm can be further extended to design 3D TMs among different symmetry classes and space groups, and different physical systems.

# 1. Introduction

Metamaterials are well-known for its novel modulation of photons, phonons, and matter waves in various applications. Enriching with the topological characteristics, it gives out a new innovative material—topological materials (TMs), which are robust to various defects[1–3]. Recently, the photonic and phononic TMs have attracted a great research interest in engineering topological phenomena on the macroscopic scale[2–12]. Related topologically protected states revealed some prospective applications[2,3,12–23].

For example, the quantum spin/valley Hall topological insulators guide the energy flux in a spin-locked transmission, which alternatively switches the one-way tunnel for the propagated waves with immunity to defects[6,8,9,24–28]. That is an ideal way to improve the effectivity of applications in opto-mechanics, current semiconductor and integrated circuit industry[13–22]. For the higher-order topological materials (HOTIs), it is characterized by an intensively localized topological phase within the lower dimensional domain such as the edges and corners[3,7,29–31]. As a pioneering example of the HOTIs, the multipole topological insulators (MTIs) can also provide a multipole moment enhanced topological phases, where the bulk dipole is vanished[3,30,31]. Together with the pseudo-spin phenomenon, a helical multipole-induced topological phase is inherited in the helical MTI[32,33]. Those topological phases in the HOTIs are robust to various defects in manufacture, and show a promising prospective in optical/acoustic subwavelength imaging, microelectronics, laser aspects[3,19,23,31,34–36].

Although the theoretical tight binding models have been developed, how to efficiently design 3D unit cells with the demanded topological behaviors is still a crucial challenge. Some typical options include tracing the featured degenerated states near the Dirac points, restricting a special band structure from the band folded mechanism, keeping an obvious Berry curvature (a quantity to topology), or realizing the maximal pseudo-spin energy fluxes in the crossing waveguide[37–44]. To calculate the topological invariants, however, it is very expensive to integrate the Berry curvature or its related terms in the whole Brillouin zone. This issue would be more pronounced for the 3D continuous TMs, which generally cover an infinite design parameter space and are more computationally expensive for analysis and design optimization.

Luckily, the theoretical breakthrough in topological quantum chemistry gives new insight into this bottleneck, from the fruitful meeting between chemistry and physics (in the real and momentum space)[45–48]. The fundamental tool is calculating the real space orbits for every band, with the aids of the elementary band representations (EBRs) or the symmetry indicators (SIs). It gives out the topology in a simple linear function of the symmetry characters at some listed point[29,30,46,48]. Successful applications of the SI method include the classification of TMs among the whole 230 space and 1651 magnetic groups, and the discovery of thousands of TMs with many uncovered for the first time[47,49,50]. Most recently, the catalogue of topological phononic materials becomes an attractive focus[4,5,11]. This inspires us to efficiently identify the topological properties of the 3D mechanical unit cells using the SI method. It is worth to note that, in the mechanical system, rigid body motions corresponding to zero energy states yield the singularity at some high symmetry point. As a result, the classic formulas derived in the quantum mechanics system need to be modified at first.

Furthermore, how to describe the 3D unit cells is an essential factor for choosing design optimization method[39,41,43,44,51]. This is because the topological invariants are discrete variables, which cannot in general effectively handled by the gradient-based algorithms. The Moving Morphable Component (MMC) method could describe 3D unit cells using only a few explicit geometry parameters, and this makes it suitable to guarantee a computationally tractable solution process of the inverse design formulation[52,53]. In general, we summarize three characteristics of a desired optimization framework of the TMs as: (i) effective to identify the topological characters for arbitrary unit cells; (ii) suitable to topological materials in different classes and different physical systems; (iii) efficient to execute the solution procedure.

In this study, we proposed a unified optimization framework for the 3D continuum TMs by combining the SI method and the MMC method. In this design framework, the helical MTIs with the helical edge and helical corner states can be effectively obtained by simultaneously constraining the modified fractional corner charge and pseudo-spin invariants. The proposed method thoughtfully modified the topological invariants for 3D elastic HOTIs according to the singularity points with zero energy, and successfully obtained optimized 3D helical MTIs in different symmetries, where the in-gap corner states are derived from the quadrupole moment. The numerical simulations of the transmission spectra and crossing waveguide applications validated the intensive corner energy and the spin-locked energy flux.

The rest of the paper is organized as follows: in Section 2, the governing equations for elastic waves are introduced. And then based on the description method of 3D elastic unit cells using the MMC method and the specialized formulas of topological invariants in Sections 3 and 4, an efficient deign paradigm is proposed for the mechanical helical MTIs. Optimized designs with different symmetries are presented in Section 6 together with the applications in a novel crossing elastic waveguide. Finally, some concluding remarks are discussed in Section 7.

## 2. The governing equations for elastic waves

In 3D elastic mechanics, the harmonic wave formulation is expressed as[54]

$$(\lambda + \mu)\nabla(\nabla \cdot \boldsymbol{u}) + \mu\nabla^2\boldsymbol{u} = -\omega^2\rho\boldsymbol{u} \tag{1}$$

where $\omega$ is the angular frequency, $\lambda$, $\mu$, and $\rho$ are the Lame's parameters and mass density, and $\boldsymbol{u} = (u, v, w)^\top$ denotes the displacement field.

Since the considered phononic crystal is periodic, the displacement field satisfies the translation condition as $\boldsymbol{u}(\boldsymbol{r} + \boldsymbol{R}) = \boldsymbol{u}(\boldsymbol{r})$ with $\boldsymbol{R}$ denoting the primitive lattice vector. According to the Bloch theorem, the harmonic elastic wave propagation can be determined by the following discretized equations

$$\begin{aligned} \mathbf{KU} &= -\omega^2\mathbf{MU} \\ \mathbf{U}(\boldsymbol{r}_0 + \boldsymbol{R})|_{\text{BC}} &= \mathbf{U}(\boldsymbol{r}_0)|_{\text{BC}}\, e^{i\boldsymbol{k}\cdot\boldsymbol{R}} \end{aligned} \tag{2}$$

Here, the matrices $\mathbf{K}$ and $\mathbf{M}$ refer to the stiffness and mass matrixes, and $\mathbf{U}$ is the eigenvector.

In order to handle the periodic constraints in the above eigenvalue problem, the standard Lagrange multiplier method is adopted[55]. By defining a Lagrange multiplier $\boldsymbol{\Lambda}$, **Eq.** (2) can be reformulated as

$$\begin{bmatrix} \mathbf{K} + \omega^2\mathbf{M} & \mathbf{N}_\text{f} \\ \mathbf{N} & \mathbf{0} \end{bmatrix} \begin{bmatrix} \mathbf{U} \\ \boldsymbol{\Lambda} \end{bmatrix} = \mathbf{0} \tag{3}$$

in which the constraint matrices $\mathbf{N}$ and $\mathbf{N}_\text{f}$ are used to homogenize the eigenvalue problem.

Now, let us decompose the eigenvector with the solution $\mathbf{U}_\text{c}$ as $\mathbf{U} = \mathbf{N}_\text{null}\mathbf{U}_\text{c} + \mathbf{U}_0$, where the matrix $\mathbf{N}_\text{null}$ and vector $\mathbf{U}_0$ belong to the null space of $\mathbf{N}$. An allowable value is $\mathbf{U}_0 = \mathbf{0}$. After left multiplying **Eq.** (3) by $\mathbf{N}_\text{nullf}^\top$ ($\mathbf{N}_\text{nullf}$ is the null space of $\mathbf{N}_\text{f}^\top$), we have the final governing equation of the 3D elastic wave as

$$\mathbf{K}_\text{c}\mathbf{U}_\text{c} = -\omega^2\mathbf{M}_\text{c}\mathbf{U}_\text{c} \tag{4}$$

where the eliminated stiffness matrix is $\mathbf{K}_\text{c} = \mathbf{N}_\text{nullf}^\top \mathbf{K} \mathbf{N}_\text{null}$, and the eliminated mass matrix is $\mathbf{M}_\text{c} = \mathbf{N}_\text{nullf}^\top \mathbf{M} \mathbf{N}_\text{null}$. Because **Eq.** (3) requires $\mathbf{N}_\text{f}\boldsymbol{\Lambda} = \mathbf{0}$, it is needless to solve for it since $\boldsymbol{\Lambda}$ is useless, and a practical choice is setting $\mathbf{N}_\text{f} = \mathbf{N}^\top$, then $\mathbf{N}_\text{nullf} = \mathbf{N}_\text{null}$.

## 3. Description of the 3D unit cells via the MMC method

Structural topology optimization has been successfully applied to inverse design various topological metamaterials[11,38–44,56]. For designing the 3D elastic topological insulators, we adopt the Moving Morphable Component (MMC) method[40,42,52,53,56], which has the advantages of the explicit geometry description and improved computational efficiency.

The building block in the MMC method is a set of morphable components, described by some geometry parameters, such as the center coordinate, length, width, and thickness. As a result, through updating those geometry parameters, every component can move, morph, merge or disappear to form the optimized structure, as shown in **Fig**. 1. In this way, the optimal parameter space will be deeply shrunk, and the solution efficiency will be significantly improved.

In our work, each 3D component (the inclusion phase) is explicitly characterized by the ellipsoid with a design variable vector $\boldsymbol{D}_i = (\boldsymbol{r}_{0i}^\mathsf{T}, \boldsymbol{L}_i^\mathsf{T}, \boldsymbol{\Phi}_i^\mathsf{T})^\mathsf{T}$, i.e., the center coordinate $\boldsymbol{r}_0 = (x_0, y_0, z_0)$, the length vector of semi-axes $\boldsymbol{L} = (L_1, L_2, L_3)$, and the Euler rotation angles $\boldsymbol{\Phi} = (\theta, \phi, \gamma)$, as shown in **Fig**. 1(a). In this manner, each MMC can be explicitly determined by only 9 design variables. Furthermore, in a unit cell, the inclusion phase is identified by the topology description function $g_i(\boldsymbol{r}, \boldsymbol{D}_i)$ for each component expressed with its covered region $\Omega_i$ as

$$g_i(\boldsymbol{r}, \boldsymbol{D}_i) = \|\boldsymbol{r}'\|_2^2 - 1 = \begin{cases} > 0 & \text{if } \boldsymbol{r} \in \Omega_i \\ = 0 & \text{if } \boldsymbol{r} \in \partial\Omega_i \\ < 0 & \text{else} \end{cases} \tag{5}$$

In **Eq**. (5), the local coordinates are determined by the global coordinates $\boldsymbol{r}$ and the rotation matrix $\boldsymbol{R}(\boldsymbol{\Phi})$ as

$$r'_i = \frac{1}{L_i} \mathrm{R}_{ij}(\boldsymbol{\Phi})(r_j - r_{0j}) \tag{6}$$

According to the symmetry requirement of the unit cells, only the MMCs in a reduced design domain need to be optimized and they can be transformed to the rest part. Furthermore, all the inclusions represented by MMCs in the design domain can be smoothed by the K-S aggregation technique [57] or the Boolean operation (adopted by this work).

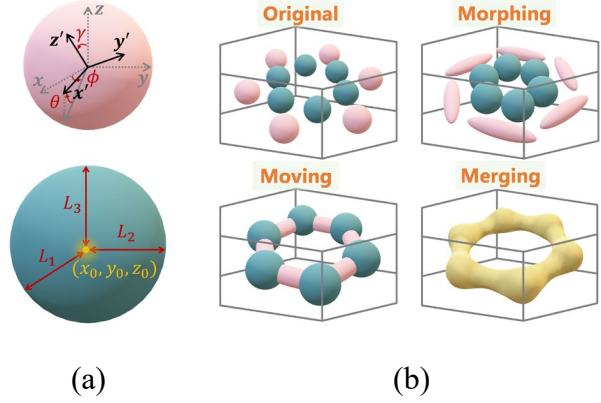

(a)  (b)

Fig. 1. An illustration of a 3D unit cell described by the MMC method. (a) The geometric description of the 3D components and (b) some representative configurations in the optimization.

## 4. The SI induced topological invariants of the helical MTIs

The helical multipole topological insulators (MTIs)[32,33], as a compound topological material, should simultaneously hold the characters (or the topological invariants) from the multipole moment and the pseudo-spin. In general, the calculation of topological invariants is computationally expensive for the continuum unit cells. Nevertheless, recent work in topological quantum chemistry reveals a rapid approach to identify topological invariants through its symmetry indicators (SIs)[4,5,46,49,50,58]. Next, we will introduce the SI method into the calculation of topological invariants, and then give out the method to design helical MTIs.

Based on the SI method, for a spinless $C_{n=3,6}$-symmetric mechanic system with the time reversal symmetry (TRS), we identify the eigenvalue $\Pi_p^{(n)}$ of the $\hat{C}_n$ rotational operator as

$$\Pi_p^{(n)} = e^{i2\pi(p-1)/n} = \langle w(\Pi)|\hat{C}_n|w(\Pi)\rangle, p \in [1,n] \qquad (7)$$

where $w(\Pi)$ denotes the z-component of the displacement at the high-symmetry point $\Pi$. The symbol $\#\Pi_p^{(n)}$ counts the number of $\Pi_p^{(n)}$ below the target band gap. Compared to the reference point $\Gamma = (0,0)$, we define the SI at $\Pi$ as $\left[\Pi_p^{(n)}\right] = \#\Pi_p^{(n)} - \#\Gamma_p^{(n)}$. At the high-symmetry points, it satisfies $\hat{C}_n \mathbf{k} = \mathbf{k} + \mathbf{G}$ with $\mathbf{G}$ denoting the reciprocal lattice vector. For the $C_3$-symmetric hexagonal unit cells, the high-symmetry points include $\Gamma$ and K in the $C_3$ symmetry, while for the $C_6$-symmetric hexagonal unit cells, they include $\Gamma$ in the $C_6$ symmetry, K in the $C_3$ symmetry, and M in the $C_2$ symmetry, respectively. In this manner, the topological classification is determined completely by the corresponding SIs, such as the fractional corner charge and the pseudo-spin invariants in the following contents.

### 4.1 The fractional corner charge invariants

For the MTIs, the fractional corner charge $Q^{(n)}$ is an effective topological invariant to determine the topological corner states[29–31]. For the 3D mechanical topological system with the TRS and $C_3$ symmetry, we propose the following formulas

$$Q^{(3)} = \left[\frac{1}{3}\left(\#K^{(3)}_{q\neq 1} - \frac{1}{2}\#\Gamma^{(3)}\right) \bmod 1\right] \times \left[(\#\Gamma^{(3)} + 1)\bmod 2\right]$$
$$Q^{(6)} = \left(\frac{1}{4}\left[\#M^{(2)}_1\right] + \frac{1}{6}\left[\#K^{(3)}_1\right]\right)\bmod 2 \tag{8}$$

Here, the red terms in the function of $Q^{(3)}$ are introduced to avoid the confused distinction of the unpaired degenerate states[6,9,10,25]. As an alternative strategy, $\#\Gamma^{(3)}$ counts the two-order degeneracy at the $\Gamma$ point, and $\#\Gamma^{(3)}/2$ identically equals the $\#\Gamma^{(3)}_2$ or $\#\Gamma^{(3)}_3$. The red modulo term in $Q^{(3)}$ guarantees the degenerate states to be in pairs (i.e., $\#\Gamma^{(3)}$ is an even number). For a visualization, the **Fig**. 2 shows that our modification successful avoids the confused distinction of degenerate states when $m = 4$. For more details, refer to **Appendix A**.

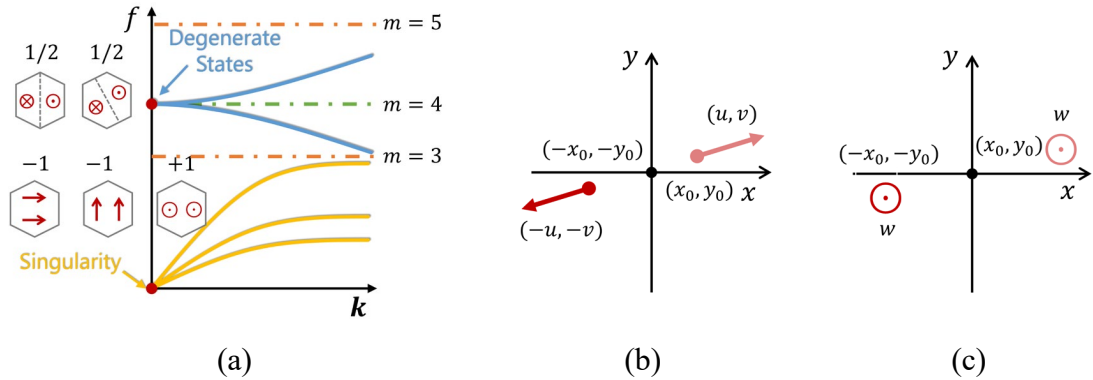

Fig. 2. The script of the modification procedure. (a) The singularity in mechanics and degenerate states in a hexagonal unit cell. (b) and (c) The vector transformation of displacement component $(u, v)$ and $w$. Here, only the degenerate states are calculated under the $\hat{C}_3$ operator, and denoted as $1/2 = (\xi + \xi^*)/2$ with its eigenvalues $\xi$ and its conjugation $\xi^*$; the other states are calculated under the $\hat{C}_2$ operator.

### 4.2 The pseudo-spin invariants

For the photonic/phononic quantum spin/valley Hall effects, the protected chiral energy flux can be well identified from the pseudo-spin vortex phenomenon and can be well quantified by the Chen-spin or $Z_2$ invariants[6,9,24,25]. In our spinless 3D mechanical system without spatial inversion symmetry, an alternative approach is adopted through tracing the (broken) Dirac cone and band inversion[9,10,25,38,40,42–44].

Practically, for the 3D $C_3$ and $C_6$ symmetric unit cells, the pseudo-spin invariants are modified as[10,29]

$$\begin{aligned} Z^{(3)} &= \text{sgn}\left(\#K_2^{(3)} - \#K_3^{(3)}\right) \\ Z^{(6)} &= \text{sgn}\left(\#\Gamma_p^{(6)} - \#\Gamma_d^{(6)} - 2\right) \end{aligned} \quad (9)$$

Here, terms $\#\Gamma_p^{(6)}$ and $\#\Gamma_d^{(6)}$ count the orbits p and d under the $\hat{C}_6$ operator for the $\Gamma$ point, respectively. Notably, the subtracted red term is introduced to correct the $\#\Gamma_p^{(6)}$ due to the singularity in the 3D elastic wave or the transverse electromagnetic wave (the singularity has the same eigenvalue as the orbit p)[5,59]. For the 3D elastic wave, the singularity relates to three translational motions, their displacement and vector transformation are shown in **Fig**. 2. This modification is based on counting all occupied bands below the target band gap, including the first three bands crossed through the singularity. Furthermore, the counting idea keeps target bands isolated from other bands, as the SI method requires. For more details, refer to **Appendix B**.

## 5. An efficient design paradigm of 3D mechanical helical MTIs

With the above topological invariants presented in Eqs. (8) and (9), we can now design the helical MTIs using explicit topology optimization method. The corresponding optimization formulation and solution process are introduced as follows.

### 5.1 Mathematical formulation

Combining the MMC-based description method and the modified formulas of topological invariants in elastic medium, optimized 3D helical MTIs can be obtained by solving the following mathematical formulation:

$$\begin{aligned} \text{find} \quad & \mathbf{D} = (\mathbf{D}_1^\top, \ldots, \mathbf{D}_N^\top, H)^\top \\ \text{max} \quad & \min(f_{\text{ref}} - \max_k f_k^m, \min_k f_k^{m+1} - f_{\text{ref}}) \\ \text{s.t.} \quad & \mathbf{K}_c \mathbf{U}_c = -\omega^2 \mathbf{M}_c \mathbf{U}_c \\ & \left(Q^{(n)}, Z^{(n)}\right) = \left(Q_{\text{ref}}^{(n)}, Z_{\text{ref}}^{(n)}\right) \\ & \mathbf{D}_{\min} \leq \mathbf{D} \leq \mathbf{D}_{\max} \end{aligned} \quad (10)$$

In the design variable vector, $D_i$ describes the $i$th component in the slab with a thickness $H$ (in the $z$-axis), as illustrated in **Fig**. 1. By denoting the eigenfrequency of the $m$th band as $f^m$, the gap width between the $m$th and $(m+1)$th bands is maximized with a target mid-frequency $f_{\text{ref}}$. The third equation in **Eq**. (10) is the governing equation for the 3D elastic waves. Since the fractional corner charge and the pseudo-spin invariants simultaneously contribute to the existence of helical corner states, the target topology invariants $\left(Q_{\text{ref}}^{(n)}, Z_{\text{ref}}^{(n)}\right)$ is introduced as a constraint. The last inequality persists the lower and upper bounds of the design variable vector.

In principle, by updating the governing equation and target topological invariants, the mathematical formulation in **Eq**. (10) can be applied for designing TMs among different symmetry classes, and different physical systems. In this work, we focused on the inverse design of 3D helical MTIs in elastic medium with $C_3$ and $C_6$ symmetries.

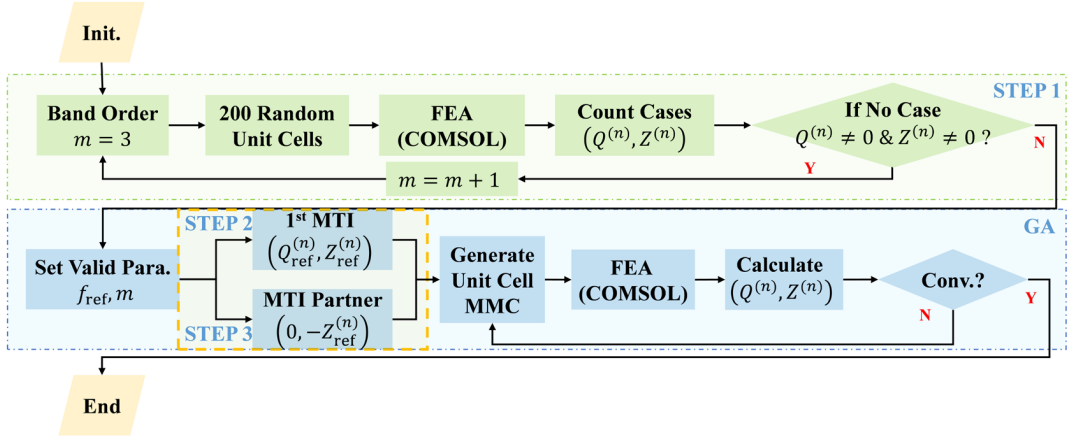

Fig. 3. The scheme of optimization for the helical MTIs.

## 5.2 Solution process

Since the topological invariants are quantized, gradient-based optimization algorithms would be ineffective for solving **Eq**. (10). Thanks to the advantage of a fewer number of design variables in the MMC method, the genetic algorithm (GA) is adopted here and the settings are presented in **Appendix C**. To be specific, the flowchart for the rational design of helical MTIs is shown in **Fig**. 3, and its solution process is summarized as follows:

- STEP 1: Initialization of the MMC method and the GA solver.

  The gap label $m$, the mid-frequency $f_{\text{ref}}$, and the nonzero topological invariants $\left(Q_{\text{ref}}^{(n)}, Z_{\text{ref}}^{(n)}\right)$ are initialized first through a trial process, starting from $m = 3$;

- STEP 2: Optimal design of the first MTI.

With the parameters determined in STEP 1, solve the mathematical programming **Eq**. (10) to obtain the first optimized MTI with the predefined invariant $\left(Q_{\text{ref}}^{(n)}, Z_{\text{ref}}^{(n)}\right)$ and mid-frequency $f_{\text{ref}}$;

- STEP 3: Optimal design of the MTI partner (if necessary).

With the desired topological invariants setting as $\left(0, -Z_{\text{ref}}^{(n)}\right)$ and the other parameters the same as STEP 2, solve **Eq**. (10) to obtain the optimized MTI partner with the inverted pseudo-spin effect.

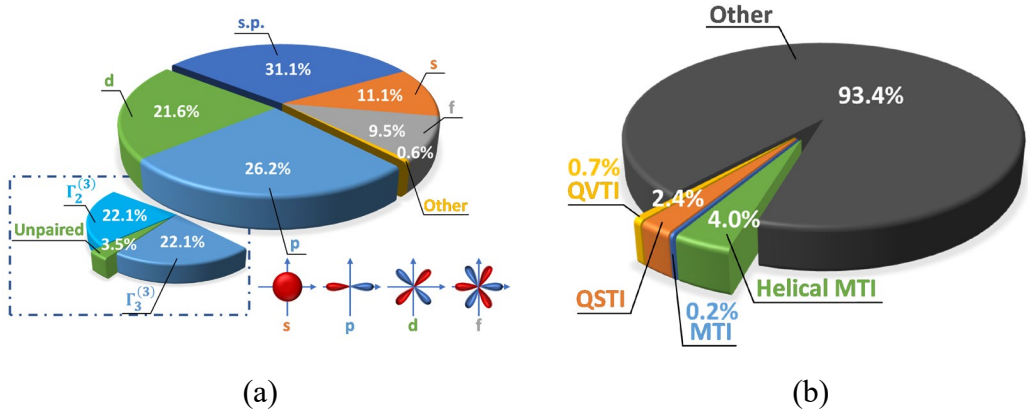

(a)             (b)

Fig. 4. The statistical charts (b) of different states at the Γ point (the partitions of p and d would be decomposed into the boxed partitions without the modification in **Eq.** (8)) and (c) of different TMs. Hint: s.p. —singularity point.

To illustrate the effectiveness of the proposed design framework, the statistical charts of the states at the Γ point (6000 samples) and of different TMs (8000 samples) are illustrated in **Fig**. 4(a) and 4(b), respectively. It can be found that, using **Eq**. (8), the states p and d are successfully identified, and they take about 21.6% and 26.2% of the whole set as shown in **Fig**. 4(a). Without the modification in **Eq**. (8), however, such states would be decomposed to $\Gamma_2^{(3)}$ state (22.1%), $\Gamma_3^{(3)}$ (22.1%), and an unpaired set of state (3.5%). This unpaired set would further make troubles for the calculation of the fractional corner charge invariant. In **Fig**. 4(b), 6.6% of 8000 samples are four typical TMs (quantum valley/spin Hall topological insulators (QVTIs/QSTIs), MTIs and helical MTIs), while the desired helical MTIs only account for 4.0%. This validates the necessity of developing inverse design paradigm for the helical MTIs.

# 6. Applications of the MMC-based design framework for 3D helical MTIs in elastic medium

In the present work, the helical MTIs are periodic in the in-plane direction and made of the basic medium EP and scattering medium Fe (materials parameters and more setup details are referred to **Appendix C**).

## 6.1 Optimal design of $C_3$-symmetric mechanical helical MTIs

Under the optimization framework, the optimized $C_3$-symmetric helical MTIs are obtained in **Fig**. 5(a). And there is a normalized bulk band gap at 0.741-1.069 between the 6th and 7th bands (colored in grey in **Fig**. 5(a)). The symmetry behaviors of the high-symmetry points are shown in the **Fig**. 5(b). There are three broken degenerate states (from the Dirac cone) at the K point below the target bandgap, while only the third one is unpaired, and implies the possibility of a pseudo-spin vortex. The phase field of this unpaired state is also inserted in **Fig**. 5(a). The corner charge and the pseudo-spin invariants are $(2/3,1)$.

In order to realize the band inversion, the corresponding MTI partner can be easily constructed by applying the spatial reversal operation, or in other words, its invariants are set as $(0,-1)$. An opposite pair of $Z^{(3)}$ invariants would produce a helical topological state from the bulk-boundary correspondence. Moreover, a pair of zero and nonzero fractional corner charges reveal the appearance of corner states[18], as shown in **Fig**. 6(a) around the normalized frequencies of 0.894 and 0.966. The latter localized corner mode is displayed in the inserted diagram.

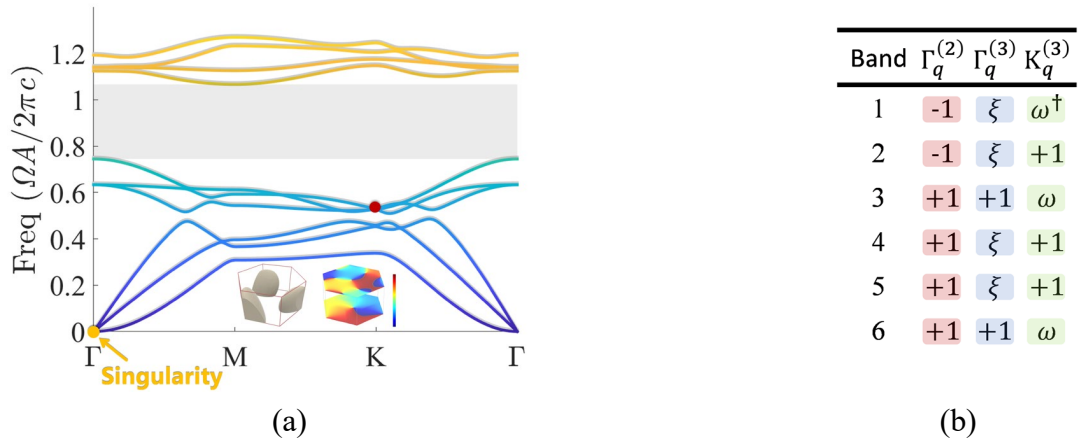

(a)                 (b)

Fig. 5. The optimized $C_3$-symmetric mechanical helical MTIs. (a) The band structure inserted with the unit cell and the phase field of the unpaired state. (b) The symmetry-

behavior-table, in which the degenerate states are tagged as $\omega = e^{i2\pi/3}$ for the K point, while for the Γ point they are tagged as $\xi$.

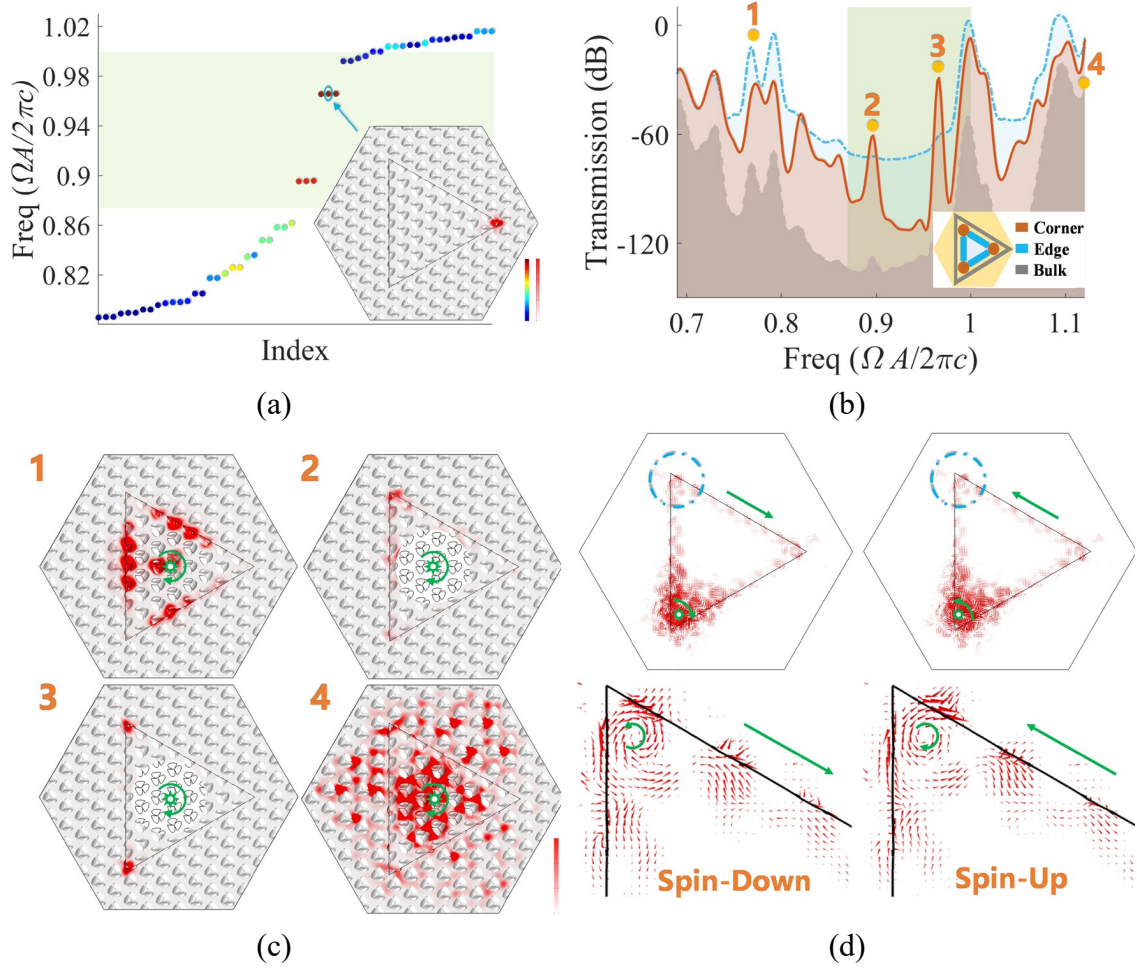

Fig. 6. Simulation results of the optimized $C_3$-symmetric helical MTIs. (a) The eigenvalue spectrum (points are colored according to the corner energy intensity) and the energy field of a corner state. (b) The transmission spectra from the probes in bulk, edge, and corner area (colored in legend). (c) Energy fields tagged in (b) at the normalized frequencies of 0.793, 0.879, 0.966, and 1.121. (d) The energy flux and their zoom-in views of helical edge states at the normalized frequency of 0.862.

Besides, the full-wave transmission is presented in **Fig**. 6(b), where energy is captured from different regions around the outer bulk, the interface edges, and the interface corners. A spin-down (clockwise) helical source is excited near the supercell's center, shown as the star in **Fig**. 6(c). The transmission reveals some edge energy peaks around the normalized frequencies of 0.793 and 1.001, and some intensively localized corner states around the normalized frequencies of 0.879 and 0.966. For a clear visualization, the corresponding bulk, edge, and corner energy fields are displayed in **Fig**. 6(c). In contrast to the edge gap around the normalized frequency range of 0.872-1.001 (colored in light-green), those in-gap corner states are derived from the quadrupole moment.

For the verification of the helical behavior, a biased helical source off the center is excited additionally, as shown in **Fig**. 6(d). The inserted arrow diagrams displayed the energy flux near their corners and edges. We found that these two supercells had significant opposite responses under different exciting helical sources (spin-up or spin-down). All their corners held a clear energy vortex (clockwise or anticlockwise). Their edge energy fluxes are locked by their exciting sources and only could flow forward or backward.

## 6.2 Optimal design of $C_6$-symmetric mechanical helical MTIs

For the optimized $C_6$-symmetric MTI pairs, as illustrated by the band structures shown in **Fig**. 7(a), band gaps are observed in the normalized frequency ranges of 1.344-1.489 (up) and 1.332-1.450 (below), respectively. Below the gap, there are four degenerate states found at the Γ points for both cases, but only the last two states formed an unpaired double Dirac cone, which features the pseudo-spin vortex. The phase fields of these unpaired states are inserted in **Fig**. 7(a), from which the band inversion is clearly displayed. The symmetry behaviors in **Fig**. 7(a) show that the corner charges and the modified pseudo-spin invariants are $(1/2,1)$ and $(0,-1)$, respectively. Specifically, the pair of opposite $Z^{(6)}$ invariants lock the energy flux by the pseudo-spin phenomenon. In contrast, the pair of zero and nonzero corner charges predict the topological corner states (according to the vanished bulk polarization in $C_6$-symmetric unit cells, these nonzero corner charges are only derived from the quadrupole moment[30]). By combining these two topological characters, the topological corner state will also have pseudo-spin behaviors and present as a helical corner state. For a verification of this helical corner state, the eigenvalue spectrum of the supercell's simulation is shown in **Fig**. 7(b), and its energy density distribution, at the normalized frequency of 1.426, is highly localized at corners.

## 6.3 Applications of the optimized helical MTIs in a crossing waveguide

As an application of the helical MTIs, a crossing waveguide (a single layer) composed of the two optimized $C_3$-symmetric helical TMIs in **Subsection** 6.1 (colored blue/yellow for the original/inversed TMIs mentioned above) is developed in **Fig**. 8(a). Since the additional pseudo-spin freedom locks the energy flux in the waveguide, two opposite transmissions would be discovered when we sequentially excited the Port 1 and Port 2. By gradually modulating the exciting frequency, the energy will spread through the center wall and induce the output corner states.

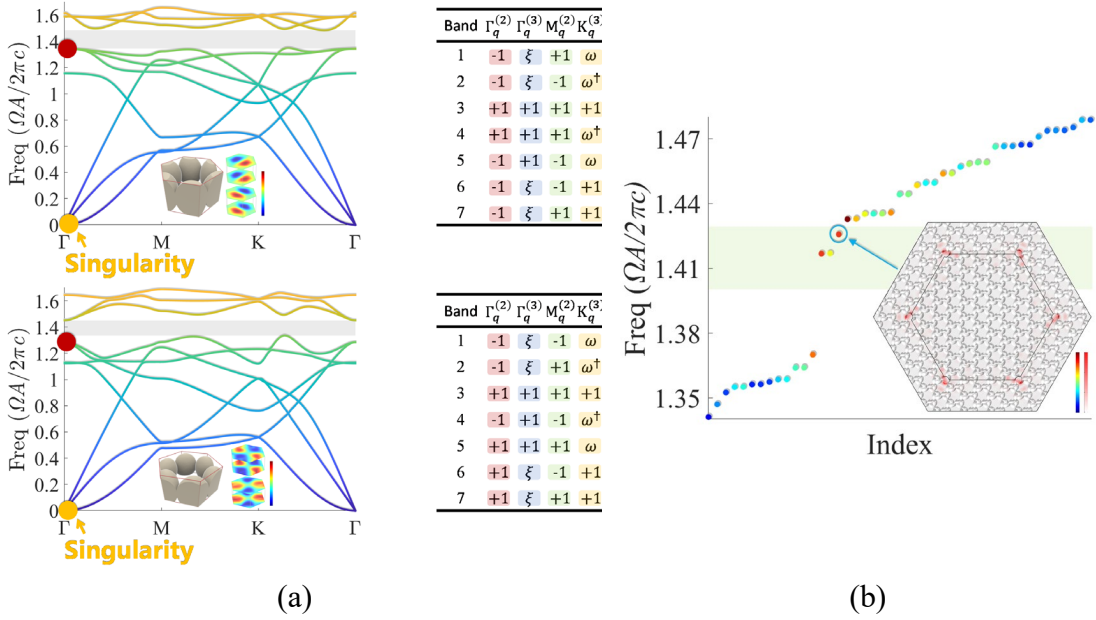

(a)  (b)

Fig. 7. Simulation results of the optimized $C_6$-symmetric helical MTIs. (a) The band structures and the symmetry-behavior-tables of the optimized MTI pairs. The inserted diagrams include the optimized unit cells and the z-directional displacement fields of the unpaired states. In those tables, the degenerate states for the K point are tagged as $\omega = e^{i2\pi/3}$, while for the Γ point they are tagged as $\xi$. (b) The eigenvalue spectrum and the inserted energy field of the corner state (points are colored according to the corner energy intensity).

The simulations in the normalized frequency range of 0.7-1.1 are processed to test the performance of the waveguide, as shown in **Fig**. 8(b). It is clear that when Port 1 is excited at the normalized frequency of 0.776, the energy only transmits to Port 2 and Port 3, yet it only transmits to Port 1 and Port 4 from Port 2. This phenomenon reveals the locked helical energy flux as expected. At the normalized frequency of 0.897, the corner states in the lower half of the waveguide are excited in both cases. Here, these states stay in the band gap of the edge states (i.e., 0.872-1.001, refer to **Appendix D** for more details**)**, and their energy only localizes at corners, and no edge states exist.

To test the working range of the one-way transmission in this waveguide, we distinguished the energy from the different ports (Port 3 or Port 4), as shown in **Figs**. 8(c) and 8(d). Here the light-green area and yellow-solid points refer to the band gap of the edge states and the states in **Fig**. 8(b). In this much wider frequency range of 0.749-0.861, the average difference between both ports is higher than 10dB. When we reverse the exciting port, the output port, which has a higher transmission, is also turned, as shown in **Fig**. 8(d). In this frequency range, the first two edge bands, as illustrated in **Appendix D**, will be excited.

Hence, these one-way transmission results from the helical edge states. Moreover, the corner states tagged with the number 2 and 4 are in the gap of the edge state but display an apparent energy concentration from the exciting source.

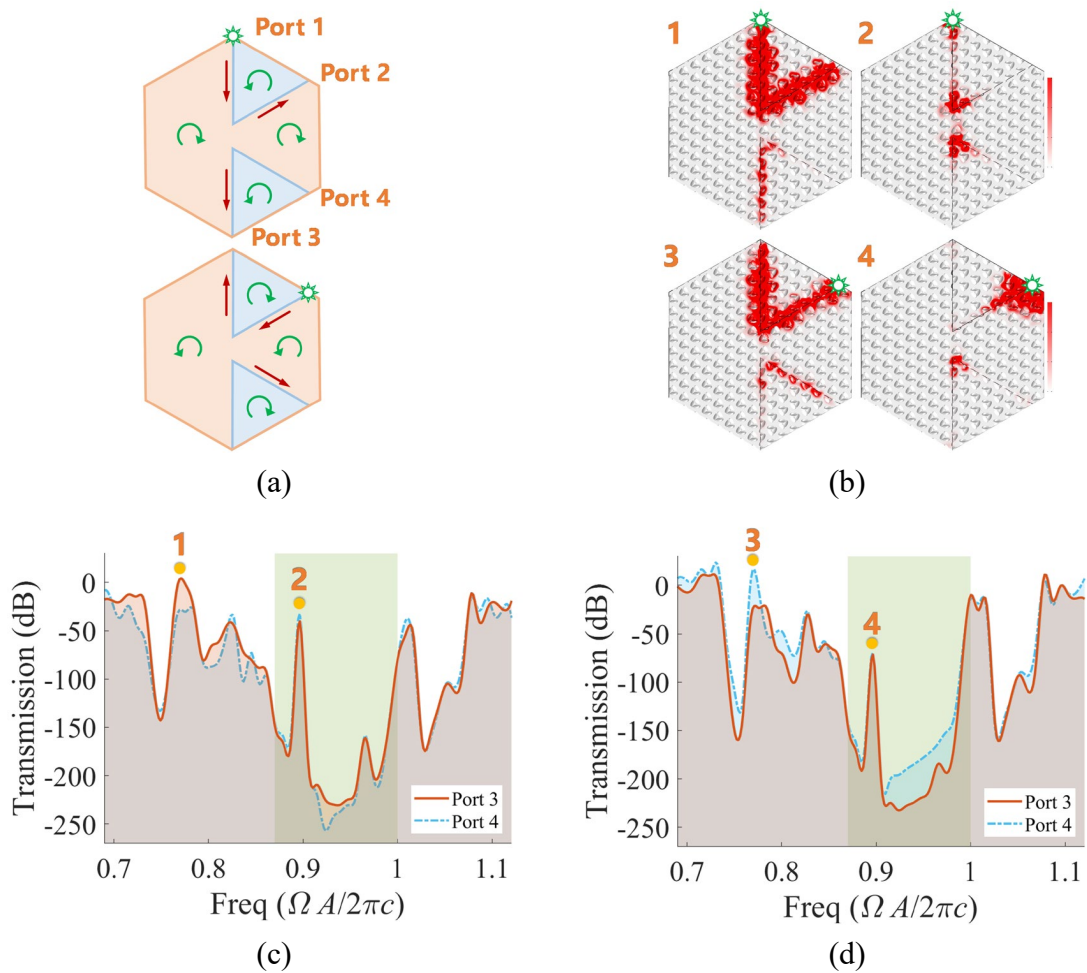

Fig. 8. The crossing waveguide made of the optimized $C_3$-symmetric helical MTIs. (a) The sketches of the waveguide and its energy fluxes in different exciting cases (the exciting line sources are tagged as stars). (b) The energy fields at the normalized frequencies of 0.776 and 0.897. The measured transmission of Ports 3 and 4 (c) from the exciting Port 1 or (d) from the exciting Port 2. Here the band gap of the edge states (light-green region) and the typical states (yellow-solid points) are colored.

## 7. Concluding remarks

In this work, we proposed an optimization framework for the inverse design of multi-functional topological materials in the 3D continuous medium. By carefully manifesting the degenerate states and singularity points in the elastic waves, the 3D helical multipole topological insulators are well-classified by the fractional corner charge and the pseudo-spin invariants. With the explicit topology optimization and the symmetry indicator methods, the proposed design paradigm has the advantages of (1) rapid classification of the 3D topological materials and (2) efficient optimization of the 3D continuum unit cells in a smaller explicit parameter space. This framework shows outstanding suitability to the 3D topological system and can also be generalized to other symmetry classes and space groups. Besides, building up a topological materials library in continuous medium would be an exciting topic for further research.

## Methods

The solid mechanic simulation is performed in the commercial software COMSOL MULTIPHYSICS. The default open surfaces are set as free boundaries. The Bloch theorem is numerical expressed by the Floquet periodic boundaries. In common, the energy in solid mechanics is consistent in distribution as the amplitude of total displacement $\|(u,v,w)\|_2^2$.

## Acknowledgements

The financial supports from the National Natural Science Foundation (11821202, 11732004, 12002073, 12002077, 12272075, 11922204), the National Key Research and Development Plan (2020YFB1709401), Dalian Talent Innovation Program (2020RQ099), the Fundamental Research Funds for the Central Universities (DUT20RC(3)020, DUT21RC(3)076), and 111 Project (B14013) are gratefully acknowledged.

## Author contributions

X. G. and Z. D. conceived the idea and initiated the project. J. L. and Z. D. established the theory. J. L and X. D. performed the numerical calculations and simulations. All the other authors contributed to the discussions of the results and the manuscript preparation.

## Declaration of competing interest

There are no conflicts to declare.

## Data availability

Data will be made available on request.

# Appendix

## Appendix A: Modification of the fractional corner charge invariant

According to the results in literature[30], the fractional corner charge invariant of the $C_3$-symmetric hexagonal unit cells is

$$Q_q'^{(3)} = \frac{1}{3}\left[K_{q\neq1}^{(3)}\right] \bmod 1 \qquad (A.1)$$

where subscript $q$ equals 2 or 3 depending on the symmetry of the constructed supercell. Due to the TRS and $C_3$ symmetry, some two-order degenerate states are protected at the $\Gamma$ point, such as states from the linear combination of the $\Gamma_2^{(3)}$ and $\Gamma_3^{(3)}$, and they are computationally expensive to identify clearly, especially for the unpaired degenerate states[6,9,10,25,28]. Instead, we termed the invariant with the number of the two-order degenerate states $\#\Gamma^{(3)}$. To be specific, the topological character of the $C_3$-symmetric hexagonal unit cell is given by

$$\bar{\chi}^{(3)} = \left(\#\Gamma^{(3)}, \#K_2^{(3)}, \#K_3^{(3)}\right) \qquad (A.2)$$

Considering **Eq.** (A.2), the modified fractional corner charge invariants and the symmetry behaviors are listed in Table A.1 for some possible cases.

Table A.1. The symmetry behaviors of the $C_3$-symmetric unit cells with TRS
(for the fractional corner charge invariants)

| $Q_{q=2}^{(3)}$ | $Q_{q=3}^{(3)}$ | $\#K_2^{(3)}$ | $\#K_3^{(3)}$ | $\#\Gamma^{(3)}$ | $\#\Gamma_2^{(3)}$ | $\#\Gamma_2^{(3)}$ |
|---|---|---|---|---|---|---|
| 1/3 | 0 | 1 | 0 | 0 | 0 | 0 |
| 0 | 1/3 | 0 | 1 | 0 | 0 | 0 |
| 0 | 2/3 | 1 | 0 | 2 | 1 | 1 |
| 0 | 0 | 1 | 1 | 2 | 1 | 1 |
| 0 | 0 | 1 | 0 | 1 | 1 | 0 |
| 0 | 0 | 0 | 1 | 1 | 0 | 1 |
| 0 | 0 | 1 | 0 | 1 | 0 | 1 |
| 0 | 0 | 0 | 1 | 1 | 1 | 0 |

In **Table** A.1, the red colored invariants $Q_q^{(3)}$ are modified from **Eq.** (A.2). This modification is derived from the fact that the degenerate states $\#\Gamma^{(3)}$ always appear in a pair; or not, it is gapless [6,9,10,25,28]. For the unpaired case, it is ambiguous to be tagged as $\Gamma_2^{(3)}$ or $\Gamma_3^{(3)}$, hence the invariants $Q_q^{(3)}$ in the last four cases cannot be solely identified by the $\bar{\chi}^{(3)}$ in **Eq**. (A.2). Therefore, when $\#\Gamma^{(3)}$ is odd, the corresponding fractional corner charge should be modified into zero with gapless band reality. When $\#\Gamma^{(3)}$ is even, number $\#\Gamma^{(3)}$ can be equivalently divided as: $\#\Gamma_2^{(3)} = \#\Gamma_3^{(3)} = \#\Gamma^{(3)}/2$.

Thus, the modified fractional corner charge invariant gives

$$Q^{(3)} = \left[\frac{1}{3}\left(\#K_{p\neq 1}^{(3)} - \frac{1}{2}\#\Gamma^{(3)}\right) \bmod 1\right] \times \left[(\#\Gamma^{(3)} + 1) \bmod 2\right] \quad (A.3)$$

where the red module term aims to avoid the unpaired two-order degenerate states.

## Appendix B: Modification of the pseudo-spin invariant

For the $C_3$ or $C_6$-symmetric unit cells with the TRS, an alternative approach to get the pseudo-spin invariants is to trace the broken (double) Dirac cones at the K or Γ point [9,10,25,38,40,42–44]. Thus, their topological characters are given by

$$\begin{aligned}\bar{\chi}^{(3)} &= \left(\#K_2^{(3)}, \#K_3^{(3)}\right) \\ \bar{\chi}^{(6)} &= \left(\#\Gamma_1^{(2)}, \#\Gamma_2^{(2)}, \#\Gamma^{(3)}\right)\end{aligned} \quad (B.1)$$

Here, $\#\Gamma^{(3)}$ counts the two-order degenerate states of the Γ point in the $C_3$ operator. The modified pseudo-spin invariants and their symmetry behaviors are listed in **Table** B.1 and **Table** B.2 for some possible cases.

Different as scaling a function[5,59], the operation of a symmetry operator $\hat{g}$ on a vector function $\boldsymbol{f}(\boldsymbol{r})$ transforms as $\hat{g}\boldsymbol{f}(\boldsymbol{r}) = \hat{R}\boldsymbol{f}(\hat{g}^{-1}\boldsymbol{r})$, where $\hat{R}$ is the rotational operator in $\hat{g}$. For the present symmetry groups ($C_3$ or $C_6$) in our paper, all group elements behave as a rotation around $z$-axis, and the transformation can be simplified as $\hat{g}\mathbf{U}(\boldsymbol{r}) = \hat{R}\mathbf{U}_\text{T}(\hat{g}^{-1}\boldsymbol{r}) + \hat{R}\mathbf{U}_\text{L}(\hat{g}^{-1}\boldsymbol{r})$, where $\mathbf{U} = \mathbf{U}_\text{T} + \mathbf{U}_\text{L} = (u, v, 0)^\top + (0,0,w)^\top$ is the displacement vector in mechanics. This decomposed equation implies $\mathbf{U}_\text{T}$ and $\mathbf{U}_\text{L}$ hold the same symmetry, except for the singular cases with displacement component $\mathbf{U}_\text{T} = \mathbf{0}$ or $\mathbf{U}_\text{L} = \mathbf{0}$.

Table B.1. The symmetry behaviors of the $C_3$-symmetric unit cells with TRS
(for the pseudo-spin invariants)

| $Z^{(3)}$ | $\#K_2^{(3)}$ | $\#K_3^{(3)}$ |
|---|---|---|
| 1 | 1 | 0 |
| 1 | 2 | 0 |
| -1 | 0 | 1 |
| 0 | 1 | 1 |

Table B.2. The symmetry behaviors of the $C_6$-symmetric unit cells with TRS
(for the pseudo-spin invariants)

| $Z^{(6)}$ | $\#\Gamma_1^{(2)}$ | $\#\Gamma_2^{(2)}$ | $\#\Gamma^{(3)}$ | Orbits |
|---|---|---|---|---|
| -1 | 2 | 0 | 2 | 2d |
| 1 | 0 | 2 | 2 | 2p |
| 0 | 2 | 2 | 4 | 2p + 2d |
| 0 | 1 | 0 | 0 | 1s |
| 0 | 0 | 1 | 0 | 1f |
| <span style="color:red">0</span> | <span style="color:red">1</span> | <span style="color:red">2</span> | <span style="color:red">2</span> | <span style="color:red">2 s.p.</span> |

In the original SI theory[29,30], the occupied bands counted in the SI method should be isolated from others, and an alternative approach is to count all bands below the target band gap. For the photonic and phononic systems, however, the first two or three bands always converge to plane waves when $|\boldsymbol{k}| \to 0$, where transverse modes produce two singularities with $\mathbf{U}_\mathrm{L} = \mathbf{0}^{5,59}$. In **Table** B.2, the red-colored data reveal the symmetry behaviors of the first three bands that always cross through the singularities around the zero energy. Hence, we defined the modified pseudo-spin invariants to overcount those singularities as

$$Z^{(3)} = \mathrm{sgn}\left(\#K_2^{(3)} - \#K_3^{(3)}\right)$$
$$Z^{(6)} = \mathrm{sgn}\left(\#\Gamma_p^{(6)} - \#\Gamma_d^{(6)} - 2\right) \quad \text{(B.2)}$$

where the red term is the modification from the singularities. For the case of photonics, the Eq. (B.2) should be further modified as the work[5]. And the terms $\#\Gamma_p^{(6)}$ and $\#\Gamma_d^{(6)}$ count the p and d states at the $\Gamma$ point.

Table C.1. Some typical optimized unit cells for the $C_3$ and the $C_6$-symmetric TMs.
(Here, the red data refers to the examples presented in our paper)

| $C_n$ | $r\{A\}$ | $L\{A\}$ | $\Phi\{deg\}$ | $H\{A\}$ | $(Q^{(n)}, Z^{(n)})$ | $f_{max}^{(m)} \sim f_{min}^{(m+1)}$ |
|---|---|---|---|---|---|---|
| $C_3$ | (0.7217,0.95,0.0275) | (0.3382,0.5073,0.1691) | (108,18,126) | 0.55 | (2/3, 1) | 0.741~1.069 |
|  | (0.4041,1.00,0.0900) | (0.3082,0.3082,0.2568) | (144,72,90) | 0.45 | (2/3, -1) | 0.772~1.000 |
| $C_6$ | (0.9238,0.90,0.3250) | (0.3682,0.2455,0.3068) | (90,108,0) | 0.65 | (1/2, 1) | 1.344~1.489 |
|  | (0.8776,0.90,0.1750) | (0.2155,0.3232,0.2693) | (90, 144, 72) | 0.50 | (0, -1) | 1.332~1.450 |
|  | (0.9584,0.80,0.1800) | (0.1766,0.4121,0.2355) | (126,126,72) | 0.60 | (1/2, 1) | 1.254~1.319 |
|  | (0.8603,0.95,0.1575) | (0.2055,0.3596,0.3082) | (108,144,90) | 0.45 | (0, -1) | 1.249~1.344 |

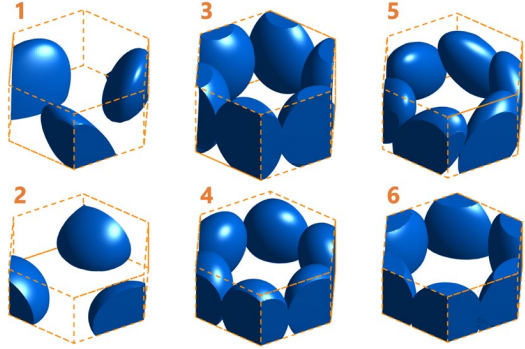

Fig. C.1. Some optimized unit cells with the $C_3$ and $C_6$ symmetries. The order refers to the row number of Table C.1.

## Appendix C: The setup of the optimization and the GA solver

For the parameters of material and optimization solver in our paper, the setup gives: the basic medium is epoxy (EP) with the elastic modulus $E_0 = 4.35\text{GPa}$, the Poisson's ratio $\nu_0 = 0.37$, and the mass density $\rho_0 = 1180\text{kg}\cdot\text{m}^{-3}$. The scattering medium is steel (Fe) with the elastic modulus $E = 200\text{GPa}$, the Poisson's ratio $\nu = 0.2$, and the mass density $\rho = 7800\text{kg}\cdot\text{m}^{-3}$. The genetic algorithm (GA) solver is set as: the population size of 100, the crossover fraction of 0.9, the migration fraction of 0.3, the elite size of 5, the objectivation tolerance of 1e-5, the stall generation limit of 15. The lattice constant is $A = |\mathbf{R}| = 1\text{m}$.

Table C.1 and Fig. C.1 list some typical optimized unit cells. The row number in Table C.1 is consistent with the order of unit cells in Fig. C.1. For the examples in the main text, we set their optimization procedure as

- For the $C_3$-symmetric helical MTIs, the broken Dirac cone appears at the K point. The first unit cell is optimized with setting $m = 6$, nonzero topological invariants (2/3,1) and no specific $f_{\text{ref}}$, which will auto-update as the mid-frequency of the target gap.

- For the $C_6$-symmetric helical MTIs, the broken Dirac cone appears at the Γ point. The first unit cell is optimized with setting $m = 7$, $\left(Q_{\text{ref}}^{(n)}, Z_{\text{ref}}^{(n)}\right) = (1/2,1)$, and $f_{\text{ref}} = 1.4$. Then the MTI partner is optimized with setting $m = 7$, $\left(Q_{\text{ref}}^{(n)}, Z_{\text{ref}}^{(n)}\right) = (0, -1)$, and $f_{\text{ref}} = 1.4$.

## Appendix D: The supercell's setup for the edge and the corner states

The setups for two example supercells are detailed as

- For the $C_3$-symmetric unit cells in the main text, the script of the truncated supercell is shown in Fig. D.1, where it provides an approach to adjust the frequency of the edge states. This truncation does not break the crystalline symmetry, and the topological edge states would not vanish. In Fig. D.1(a) and D.1(b), the truncation is set as $T = 1/4 \times 2A/\sqrt{3}$, and the edge gap is between 0.872-1.001. In Fig. D.1(c) and (d), the eigenvalue spectrum and the crossing waveguide are simulated with the truncation $T = 1 \times 2A/\sqrt{3}$. The light-blue areas in Fig. D.1 refer to the supercell's structures in the main text.

- For the $C_6$-symmetric unit cell, the inner interface between the unit cell pairs can be alternatively truncated as Fig. D.2. In Fig. D.2 (a) and D.2(b), the gap of the edge states is found between 1.398-1.425 with the truncation $T = 1/2 \times A$. The supercell's script for the eigenvalue spectrum is displayed in Fig. D.2(c) with the truncation $T = 1 \times 2A/\sqrt{3}$. In Fig. D.2(c), except for the inner hexagonal interface, six base medium cylinders with a diameter of $0.4A$ are added to adjust the supercell's corners. The light-blue areas in Fig. D.2 refer to the supercell's structures in the main text.

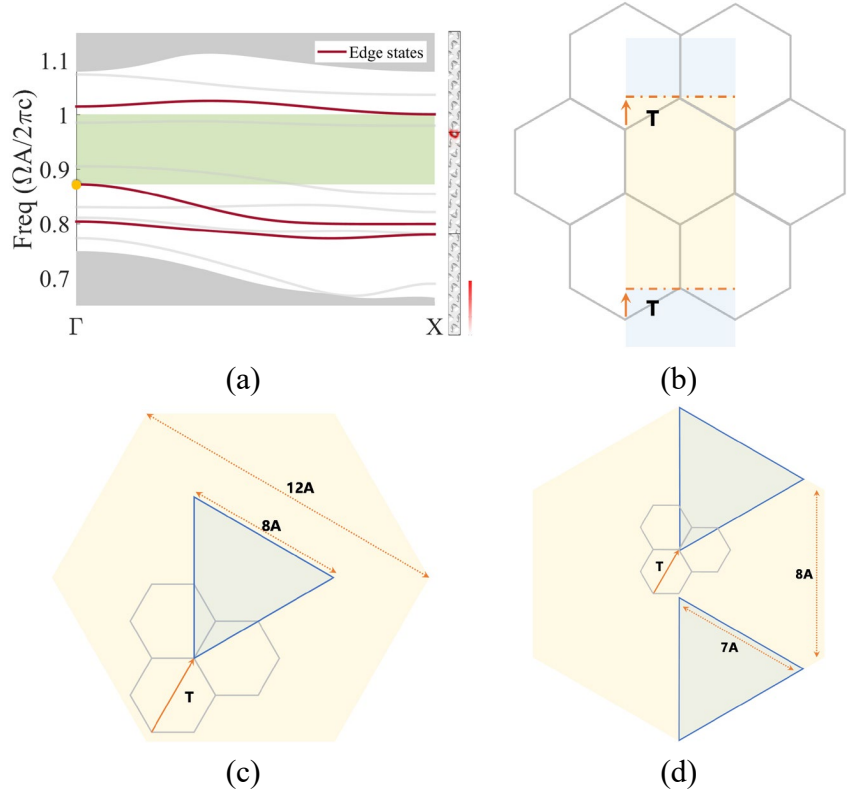

Fig. D.1. The supercell's scripts of the $C_3$-symmetric unit cell. (a) The band of the edge state (light-grey band belongs to the lower interface counterpart), and (b) the script of the ribbon-shaped supercell in (a). The supercell's script (c) for the eigenvalue spectrum and (d) for the simulation of the crossing waveguide.

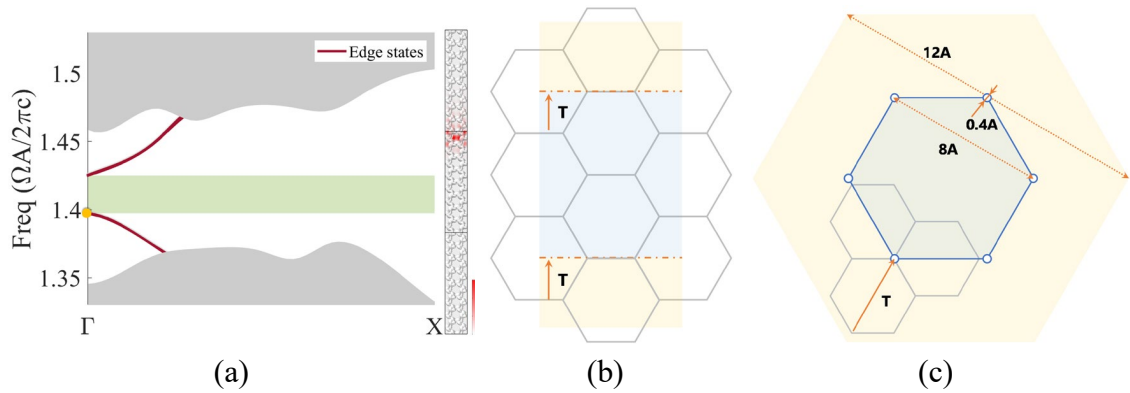

Fig. D.2. The supercell's scripts for the $C_6$-symmetric unit cell. (a) The band of the edge states, (b) the script of the ribbon-shaped supercell in (a), and (c) the supercell's script for the simulation of the crossing waveguide.

# Reference

1. Hasan, M. Z. & Kane, C. L. Colloquium : topological insulators. *Rev Mod Phys* **82**, 3045 (2010).

2. Lu, L., Joannopoulos, J. D. & Soljačić, M. Topological photonics. *Nat Photon* **8**, 821–829 (2014).

3. Xie, B. *et al.* Higher-order band topology. *Nat Rev Phy* **3**, 520–532 (2021).

4. Xu, Y. *et al.* Catalogue of topological phonon materials. *ArXiv* (2022).

5. Christensen, T., Po, H. C., Joannopoulos, J. D. & Soljačić, M. Location and topology of the fundamental gap in photonic crystals. *Phys Rev X* **12**, 21066 (2022).

6. Ma, T. & Shvets, G. All-Si valley-Hall photonic topological insulator. *New J Phys* **18**, 025012 (2016).

7. Wu, Q., Chen, H., Li, X. & Huang, G. In-plane second-order topologically protected states in elastic Kagome lattices. *Phys Rev Appl* **14**, 014084 (2020).

8. Süsstrunk, R. & Huber, S. D. Observation of phononic helical edge states in a mechanical topological insulator. *Science (1979)* **349**, 47–50 (2015).

9. Wu, L. H. & Hu, X. Scheme for achieving a topological photonic crystal by using dielectric material. *Phys Rev Lett* **114**, 223901 (2015).

10. Košata, J. & Zilberberg, O. Second-order topological modes in two-dimensional continuous media. *Phys Rev Res* **3**, L032029 (2021).

11. Peano, V., Sapper, F. & Marquardt, F. Rapid exploration of topological band structures using deep learning. *Phys Rev X* **11**, 21052 (2021).

12. Long, Y., Ren, J. & Chen, H. Intrinsic spin of elastic waves. *Proc Natl Acad Sci* **115**, 9951–9955 (2018).

13. Zhang, Z. *et al.* Directional acoustic antennas based on valley-Hall topological insulators. *Adv Mater* **30**, 1803229 (2018).

14. Zhang, Z. *et al.* Topological acoustic delay line. *Phys Rev Appl* **9**, 34032 (2018).

15. Nii, Y. & Onose, Y. Microwave impedance microscopy imaging of acoustic topological edge mode on a patterned substrate. *ArXiv* (2022).

16. Kumar, A. *et al.* Topological sensor on a silicon chip. *Appl Phys Lett* **121**, 11101 (2022).

17. Hafezi, M., Demler, E. A., Lukin, M. D. & Taylor, J. M. Robust optical delay lines with topological protection. *Nat Phys* **7**, 907–912 (2011).

18. Gong, Y. *et al.* Topological insulator laser using valley-Hall photonic crystals. *ACS Photon* **7**, 2089–2097 (2020).

19. Bandres, M. A. *et al.* Topological insulator laser: experiments. *Science (1979)* **359**, eaar4005 (2018).

20. Yu, S. Y. *et al.* Elastic pseudospin transport for integratable topological phononic circuits. *Nat Commun* **9**, 3072 (2018).

21. Ma, T. X. *et al.* Flexural wave energy harvesting by the topological interface state of a phononic crystal beam. *Extreme Mech Lett* **50**, 101578 (2022).

22. Zhou, W. *et al.* Actively controllable topological phase transition in homogeneous piezoelectric rod system. *J Mech Phys Solids* **137**, 103824 (2020).

23. Zhang, Z. *et al.* Deep-subwavelength holey acoustic second-order topological insulators. *Adv Mater* **31**, 1904682 (2019).

24. Chen, Y., Liu, X. & Hu, G. Topological phase transition in mechanical honeycomb lattice. *J Mech Phys Solids* **122**, 54–68 (2019).

25. Xu, L. *et al.* Accidental degeneracy in photonic bands and topological phase transitions in two-dimensional core-shell dielectric photonic crystals. *Opt Express* **24**, 18059 (2016).

26. Kang, Y. *et al.* Pseudo-spin–valley coupled edge states in a photonic topological insulator. *Nat Commun* **9**, 3029 (2018).

27. Mousavi, S. H., Khanikaev, A. B. & Wang, Z. Topologically protected elastic waves in phononic metamaterials. *Nat Commun* **6**, 8682 (2015).

28. Zhang, Z. *et al.* Topological creation of acoustic pseudospin multipoles in a flow-free symmetry-broken metamaterial lattice. *Phys Rev Lett* **118**, 84303 (2017).

29. Schindler, F. *et al.* Fractional corner charges in spin-orbit coupled crystals. *Phys Rev Res* **1**, 33074 (2019).

30. Benalcazar, W. A., Li, T. & Hughes, T. L. Quantization of fractional corner charge in Cn-symmetric higher-order topological crystalline insulators. *Phys Rev B* **99**, 245151–13 (2019).

31. Benalcazar, W. A., Bernevig, B. A. & Hughes, T. L. Quantized electric multipole insulators. *Science (1979)* **357**, 61–66 (2017).

32. Yang, Z. Z. *et al.* Helical higher-order topological states in an acoustic crystalline insulator. *Phys Rev Lett* **125**, 255502 (2020).

33. Liu, F., Deng, H. Y. & Wakabayashi, K. Helical topological edge states in a quadrupole phase. *Phys Rev Lett* **122**, 86804 (2019).

34. Zhang, X. *et al.* Second-order topology and multidimensional topological transitions in sonic crystals. *Nat Phys* **15**, 582–588 (2019).

35. Fan, H. *et al*. Elastic higher-order topological insulator with topologically protected corner states. *Phys Rev Lett* **122**, 204301 (2019).

36. Kim, M., Jacob, Z. & Rho, J. Recent advances in 2D, 3D and higher-order topological photonics. *Light Sci Appl* **9**, 130 (2020).

37. Lu, Y. & Park, H. S. Double Dirac cones and topologically nontrivial phonons for continuous square symmetric C4(v) and C2(v) unit cells. *Phys Rev B* **103**, 64308 (2021).

38. Zhu, X. *et al.* Topological transitions in continuously deformed photonic crystals. *Phys Rev B* **97**, 85148 (2018).

39. Chen, Y. *et al.*. Inverse design of higher-order photonic topological insulators. *Phys Rev Res* **2**, 23115 (2020).

40. Du, Z., Chen, H. & Huang, G. Optimal quantum valley Hall insulators by rationally engineering Berry curvature and band structure. *J Mech Phys Solids* **135**, 103784 (2020).

41. Dong, H. W. *et al.* Customizing acoustic Dirac cones and topological insulators in square lattices by topology optimization. *J Sound Vib* **493**, 115687 (2021).

42. Luo, J. *et al.* Moving morphable components-based inverse design formulation for quantum valley/spin Hall insulators. *Extreme Mech Lett* **45**, 101276 (2021).

43. Nanthakumar, S. S. *et al.* Inverse design of quantum spin Hall-based phononic topological insulators. *J Mech Phys Solids* **125**, 550–571 (2019).

44. Christiansen, R. E., Wang, F., Sigmund, O. & Stobbe, S. Designing photonic topological insulators with quantum-spin-Hall edge states using topology optimization. *Nanophotonics* **8**, 1363–1369 (2019).

45. Hoffmann, R. How chemistry and physics meet in the solid state. *Angew Chem Int Ed Engl* **26**, 846–878 (1987).

46. Bradlyn, B. *et al.* Topological quantum chemistry. *Nature* **547**, 298–305 (2017).

47. Po, H. C., Vishwanath, A. & Watanabe, H. Symmetry-based indicators of band topology in the 230 space groups. *Nat Commun* **8**, 50 (2017).

48. Fu, L. & Kane, C. L. Topological insulators with inversion symmetry. *Phys Rev B* **76**, 45302 (2007).

49. Watanabe, H., Po, H. C. & Vishwanath, A. Structure and topology of band structures in the 1651 magnetic space groups. *Sci Adv* **4**, eaat8685 (2018).

50. Tang, F., Po, H. C., Vishwanath, A. & Wan, X. Topological materials discovery by large-order symmetry indicators. *Sci Adv* **5**, eaau8725 (2019).

51. Bendsøe, M. P. & Sigmund, O. *Topology Optimization: Theory, Methods, And Applications*. (Springer Berlin Heidelberg, 2004).

52. Guo, X., Zhang, W. & Zhong, W. Doing topology optimization explicitly and geometrically-a new moving morphable components based framework. *J Appl Mech* **81**, 81009 (2014).

53. Du, Z. *et al.* An efficient and easy-to-extend Matlab code of the moving morphable component (MMC) method for three-dimensional topology optimization. *Struct Multidiscip Optim* **65**, 158 (2022).

54. J.D. Achenbach. *Wave Propagation In Elastic Solids*. (Elsevier Science Publishing Co Inc, 1973).

55. Strang, G. *Introduction to Linear Algebra*. (Wellesley Cambridge Press, 1993).

56. Luo, J. *et al.* Multi-class, multi-functional design of photonic topological insulators by rational symmetry-indicators engineering. *Nanophotonics* **10**, 4523–4531 (2021).

57. Kreisselmeier, G. & Steinhauser, R. Systematic control design by optimizing a vector performance index. *Computer Aided Design of Control Systems* 113–117 (1980).

58. Po, H. C., Vishwanath, A. & Watanabe, H. Complete theory of symmetry-based indicators of band topology. *Nat Commun* **8**, 50 (2017).

59. Watanabe, H. & Lu, L. Space group theory of photonic bands. *Phys Rev Lett* **121**, 263903 (2018).